\documentclass[11pt]{article}
\usepackage{fullpage}
\usepackage{graphicx}
\usepackage{amssymb}
\usepackage{epstopdf}
\usepackage{amsmath}
\title{Entry \#100742: Degenerate Rayleigh-Plateau instability in a magnetically annealed colloidal dispersion}
\author{James W. Swan, Yifei Liu and Eric M. Furst\\Department of Chemical and Biomolecular Engineering, University of Delaware}

\date{}

\begin{document}

\maketitle

\begin{abstract}
This fluid dynamics video depicts the evolution of a suspension of paramagnetic colloids under the influence of a uniform, pulsed magnetic field.  At low pulse frequencies, the suspension condenses into columns which decompose via a Rayleigh-Plateau instability.  At high pulse frequencies, the suspension forms a kinetically arrested, system spanning network.  We demonstrate the degeneration of the Rayleigh-Plateau instability with increasing pulse frequency. 
\end{abstract}

Super-paramagnetic colloids (525 nm in radius, Invitrogen MyOne), forming a monolayer with concentration $ 70\% $ by area on a microscope slide, are imaged via dark field microscopy.  In the video micrographs, lighter areas indicate particle-rich regions, where the illumination is scattered, while black regions are particle-free.  The microscope slide is placed in a Helmholtz coil that generates a uniform magnetic field with field lines oriented vertically in the micrograph.  The magnetic field with strength $ 1500 $ A/m induces a dipolar interaction between the particles that is many times stronger than the thermal energy scale $ kT $.  Thermodynamics predicts that the suspension phase separates into a particle-poor fluid and a body centered tetragonal crystal under these conditions (A.P. Hynninen and M. Dijkstra, Phys. Rev. Lett., 94:138303, 2005).  Instead, however, a steady magnetic field of this magnitude leaves the suspension kinetically arrested.

To subvert this arrested state, we pulse the magnetic field on and then off, periodically in time.  Pulse frequencies, $\omega$, which are high relative to the relaxation rate of the suspension, are indistinguishable from steady state.  Lower pulse frequencies lead to a increased particle mobility and a microstructure that evolves towards the state predicted for equilibrium.  The transition from disordered monolayer to condensed phase proceeds due to a remarkable Rayleigh-Plateau instability when the pulse frequency is low.

In the video we depict the evolution of the suspension driven by magnetic fields with four different pulse frequencies.  Beginning with the pulse frequency, $ 0.66 $ Hz, we observe that the suspension organizes into dense columns or threads parallel with the magnetic field lines.  These columns are unstable and their surface deforms sinusoidally.  This perturbation grows until the columns break up into interacting magnetic droplets.  This is the classical Rayleigh-Plateau instability.

At a pulse frequency of $ 1 $ Hz, similar columnar structures form initially and rapidly decompose into elongated droplets.  Because the field is turned off for a shorter duration when $ \omega = 1 $ Hz than $ 0.66 $ Hz, the particle rich domains can be thought of as relaxing more slowly -- or being more viscous.  The theory of Tomotika (Proc. Royal Soc. Lon. A. 150(870):332Ð337, 1935) suggests that the most unstable wavelength of perturbation in the Rayleigh-Plateau instability grows with the viscosity contrast of the fluid columns relative to the solvent in which it is immersed.  The instability will appear to degenerate with increasing frequency because the most unstable wavelength is increasing.

We illustrate this degeneration through evolution of the suspension under $ 5 $ Hz and $ 10 $ Hz pulses.  Here, the particles are largely arrested and only the coarsest transitions in structure are observed.  Rich illumination from the particulate phase indicates the presence of scattering interfaces and a low local particle density.  In these cases, the suspension relaxes so slowly -- or is so viscous -- that the Rayleigh-Plateau instability goes unobserved.

Finally, these four experiments are compared side by side.  The difference in the kinematics is striking.  Pulsing the magnetic field leads to orders of magnitude reduction in the time scales necessary for the suspension to condense to its equilibrium state.  The frequency of pulsing affects the route by which that condensed state is reached.  While increasing pulse frequency and thus decreasing particle mobility, the Rayleigh-Plateau instability degenerates from its classical portrayal -- threads become droplets -- to a slowly evolving, kinetically arrested mass.  Such a process taking advantage of the Rayleigh-Plateau instability could be key for advancing the directed self-assembly of ordered phases from nano-particles (Swan et al., Proc. Nat. Acad. Sci. USA, 109:16023--16028, 2012).

\end{document}